\documentclass[12pt]{article}
\usepackage{amsmath,amssymb}
\usepackage{graphicx}
\begin{document}

\title{Numerical Study of Nearly Singular Solutions of the 3-D 
Incompressible Euler Equations}
\author{Thomas Y. Hou\thanks{Applied and Comput. Math, 217-50, Caltech, Pasadena, 
CA 91125. Email: hou@acm.caltech.edu, and LSEC, 
Academy of Mathematics and Systems Sciences, Chinese Academy of Sciences,
Beijing 100080, China.}
\and Ruo Li\thanks{Applied and Comput. Math., Caltech, Pasadena, CA 91125, and
LMAM\&School of Mathematical Sciences, Peking University, Beijing, China.
Email: rli@acm.caltech.edu.}}

\maketitle

\begin{abstract}
In this paper, we perform a careful numerical study of
nearly singular solutions of the 3D incompressible Euler 
equations with smooth initial data. We consider the 
interaction of two perturbed antiparallel vortex tubes 
which was previously investigated by Kerr in \cite{Kerr93,Kerr05}. 
In our numerical study, we use both the pseudo-spectral 
method with the 2/3 dealiasing rule and the pseudo-spectral method 
with a high order Fourier smoothing. Moreover, we perform a 
careful resolution study with grid points as large as 
$1536\times1024\times 3072$ to demonstrate the convergence 
of both numerical methods. Our computational results show
that the maximum vorticity does not grow faster than doubly 
exponential in time while the velocity field remains bounded
up to $T=19$, beyond the singularity time $T=18.7$ reported
by Kerr in \cite{Kerr93,Kerr05}. The local geometric regularity
of vortex lines near the region of maximum vorticity seems 
to play an important role in depleting the nonlinear
vortex stretching dynamically.

\end{abstract}

\section{Introduction}

The question of whether the solution of the 3D incompressible
Euler equations can develop a finite time singularity from a 
smooth initial condition is one of the most challenging problems.
A major difficulty in obtaining the global regularity of the
3D Euler equations is due to the presence of the vortex
stretching, which is formally quadratic in vorticity. 
There have been many computational efforts in searching for 
finite time singularities of the 3D Euler and Navier-Stokes 
equations, see e.g. 
\cite{Chorin82,PS90,KH89,GS91,SMO93,Kerr93,Caf93,BP94,FZG95,Pelz98,GMG98,Kerr05}. 
Of particular interest is the numerical study of 
the interaction of two perturbed antiparallel vortex tubes by
Kerr \cite{Kerr93,Kerr05}, in which a finite time blowup of
the 3D Euler equations was reported. There has been a lot of 
interests in studying the interaction of two perturbed antiparallel 
vortex tubes in the late eighties and early nineties because of the
vortex reconnection phenomena observed for the Navier-Stokes 
equations. While most studies indicated only exponential growth 
in the maximum vorticity \cite{PS90,AG89,BPZ92,KH89,MH89,SMO93},
the work of Kerr and Hussain in \cite{KH89} suggested a finite 
time blow-up in the infinite Reynolds number limit, which 
motivated Kerr's Euler computations mentioned above.

There has been some interesting development in the theoretical understanding
of the 3D incompressible Euler equations. It has been shown that the local 
geometric regularity of vortex lines can play an important role in depleting 
nonlinear vortex stretching \cite{Const94,CFM96,DHY05a,DHY05b}. In particular, 
the recent results obtained by Deng, Hou, and Yu \cite{DHY05a,DHY05b} 
show that geometric regularity of vortex lines, even in an extremely 
localized region containing the maximum vorticity, can lead to depletion of 
nonlinear vortex stretching, thus avoiding finite time singularity formation 
of the 3D Euler equations. 

In a recent paper \cite{HL06}, we have performed well-resolved computations 
of the 3D incompressible Euler equations using the same initial condition 
as the one used by Kerr in \cite{Kerr93}. In our computations, we use
a pseudo-spectral method with a very high order Fourier smoothing to 
discretise the 3D incompressible Euler equations. The time integration 
is performed using the classical fourth order Runge-Kutta method with 
adaptive time stepping to satisfy the CFL stability condition. We use up 
to $1536\times 1024\times 3072$ space resolution to resolve the nearly 
singular behavior of the 3D Euler equations. Our computational results 
demonstrate that the maximum vorticity does not grow faster than 
doubly exponential in time, up to $t=19$, beyond the singularity time 
$t=18.7$ predicted by Kerr's computations \cite{Kerr93,Kerr05}. Moreover, 
we show that the velocity field, the enstrophy, and enstrophy production 
rate remain bounded throughout the computations. This is in contrast
to Kerr's computations in which the vorticity blows up like
$O((T-t)^{-1})$ and the velocity field blows up like $O((T-t)^{-1/2})$. 
The vortex lines near the region of the maximum vorticity are found to
be relatively smooth. With the velocity field being bounded, the non-blowup
result of Deng-Hou-Yu \cite{DHY05a,DHY05b} can be applied, which implies that 
there is no blowup of the Euler equations up to $T=19$. The local geometric 
regularity of the vortex lines near the region of maximum vorticity 
seems to play an important role in the dynamic depletion of vortex 
stretching.

The purpose of this paper is to perform a systematic convergence study 
using two different numerical methods to further validate the 
computational results obtained in \cite{HL06}. These two methods 
are the pseudo-spectral method with the 2/3 dealiasing rule and the 
pseudo-spectral method with a high order Fourier smoothing. For the 
3D Euler equations with periodic boundary conditions, the pseudo-spectral 
method with the 2/3 dealiasing rule has been used widely in the 
computational fluid dynamics community. This method has the advantage 
of removing the aliasing errors completely. On the other hand, when 
the solution is nearly singular, the decay of the Fourier spectrum 
is very slow. The abrupt cut-off of the last 1/3 of its Fourier modes 
could generate significant oscillations due to the Gibbs phenomenon.
In our computational study, we find that the pseudo-spectral method 
with a high order Fourier smoothing can alleviate this difficulty by 
applying a smooth cut-off at high frequency modes. Moreover, we find 
that by using a high order smoothing, we can retain more effective 
Fourier modes than the 2/3 dealiasing rule. This gives a better 
convergence property. To demonstrate the convergence of both methods, 
we perform a careful resolution study, both in the physical space 
and spectrum space. Our extensive convergence study shows that both 
numerical methods converge to the same solution under mesh refinement. 
Moreover, we show that the pseudo-spectral method with a high order 
Fourier smoothing offers better accuracy than the pseudo-spectral 
method with the 2/3 dealiasing rule. 

To understand the differences between our computational results and 
those obtained by Kerr in \cite{Kerr93}, we need to make some
comparison between Kerr's computations \cite{Kerr93} and our 
computations. In Kerr's computations, he used a pseudo-spectral 
discretization with the 2/3 dealiasing rule in the $x$ and $y$ directions, 
and a Chebyshev discretization in the $z$-direction with resolution 
of order $512\times 256 \times 192$. In order to prepare the initial 
data that can be used for the Chebyshev polynomials, Kerr performed 
some interpolation and used extra filtering. As noted by Kerr 
\cite{Kerr93} (see the top paragraph of page 1729), ``An effect of 
the initial filter upon the vorticity contours at $t=0$ is a long 
tail in Fig. 2(a)'' (see also Figure \ref{fig.vort-init1} of this 
paper). Such ``a long tail'' is clearly a numerical artifact. In 
comparison, since we use pseudo-spectral approximations in 
all three directions, there is no need to perform interpolation 
or use extra filtering as was done in \cite{Kerr93}. Our initial 
vorticity contours are essentially symmetric (see Figure 
\ref{fig.vort-init}). There is no such ``a long tail''
in our initial vorticity contours. It seems reasonable to expect
that the ``long tail'' in Kerr's discrete initial condition could
affect his numerical solution at later times.

A more important difference between Kerr's computations and our
computations is the difference between his numerical resolution 
and ours. From the numerical results presented at $t=15$ and 
$t=17$ in \cite{Kerr93}, one can observe noticeable oscillations 
in the vorticity contours (see Figure 4 of \cite{Kerr93} or Figure 
\ref{fig.kerr-scaled_local_struc} of this paper). By $t=17$, 
the two vortex tubes have effectively turned into two thin 
vortex sheets which roll up at the left edge (see Figures
\ref{fig.local_struc_3d_17} and \ref{fig.local_struc_3d} of 
this paper). The rolled up portion of the vortex sheet 
travels backward in time and moves away from the dividing 
plane (the $x-y$ plane).  With only 192 Chebyshev grid points 
along the $z$-direction in Kerr's computations, there are not 
enough grid points to resolve the rolled up portion of the 
vortex sheet, which is some distance away from the dividing 
plane. The lack of resolution along the $z$-direction plus 
the Gibbs phenomenon due to the use of the 2/3 dealiasing 
rule in the $x$ and $y$ directions may contribute to the 
oscillations observed in Kerr's computations. In comparison, 
we have 3072 grid points along the $z$-direction, which 
provide about 16 grid points across the singular layer at 
$t=18$, and about 8 grid points at $t=19$ \cite{HL06}. It is 
also worth mentioning that Kerr has only about 100 effective 
Fourier modes in the $x$ and $y$ directions (see Figure 18 of 
\cite{Kerr93}), while we have about 1300 effective Fourier 
modes in $|k|$ (see Figures \ref{fig.enstrophy_spec_1}
and \ref{fig.energy_spec_1} of this paper). The difference 
between our resolutions is clearly significant.

It is worth noting that the computations for $t \le 17$, which Kerr 
used as the primary evidence for a singularity, is still far from the 
predicted singularity time, $T=18.7$. With the asymptotic scaling 
parameter being $T-t = 1.7$, the error in the singularity fitting 
could be of order one. In order to justify the predicted asymptotic
behavior of vorticity and velocity blowup, one needs to perform
well-resolved computations much closer to the predicted singularity
time. As our computations demonstrate, the alleged singularity 
scaling, $\| \vec{\omega} \|_\infty \approx c/(T-t)$, does not persist
in time (here $\vec{\omega}$ is vorticity). If we take $T=18.7$, 
as suggested in \cite{Kerr93}, the
scaling constant, $c$, does not remain constant as $t \rightarrow T$.
In fact, we find that $c$ rapidly decays to zero as $t \rightarrow T$
(see Figure \ref{fig.scaling_const} of this paper).

The rest of this paper is organized as follows. We describe the set-up of 
the problem in Section 2. In Section 3, we perform a systematic
convergence study of the two numerical methods. We describe our numerical 
results in detail and compare them with the previous results obtained 
in \cite{Kerr93,Kerr05} in Section 4. Some concluding remarks are made 
in Section 5.

\section{The set-up of the problem}

The 3D incompressible Euler equations in the vorticity stream function
formulation are given as follows:
\begin{eqnarray}\label{3deuler}
\vec{\omega}_t+(\vec{u}\cdot\nabla) \vec{\omega} & = & \nabla 
\vec{u} \cdot \vec{\omega}, \\
- \bigtriangleup \vec{ \psi} &= & \vec{\omega}, \quad
\vec{u} = \nabla \times \vec{\psi},
\end{eqnarray} 
with initial condition $\vec{\omega}\mid_{t=0} =  \vec{\omega}_{0}$,
where $\vec{u}$ is velocity, $\vec{\omega}$ is vorticity, and 
$\vec{\psi}$ is stream function. Vorticity is related to velocity 
by $\vec{\omega} = \nabla \times \vec{u}$. The incompressibility
implies that 
\[
\nabla \cdot \vec{u} = \nabla \cdot \vec{\omega} = \nabla \cdot \vec{\psi} = 0.
\]
We consider periodic boundary 
conditions with period $4 \pi$ in all three directions. 

We study the interaction of two perturbed antiparallel vortex tubes using 
the same initial condition as that of Kerr (see Section III of \cite{Kerr93}). 
Following \cite{Kerr93}, we call the $x$-$y$ plane as the ``dividing plane'' and 
the $x$-$z$ plane as the ``symmetry plane''. There is one vortex tube above and 
below the dividing plane respectively. The term ``antiparallel'' refers to
the anti-symmetry of the vorticity with respect to the dividing plane
in the following sense: $\vec{\omega}(x,y,z) = -\vec{ \omega}(x,y,-z)$. 
Moreover, with respect to the symmetry plane, the vorticity is symmetric in 
its $y$ component and anti-symmetric in its $x$ and $z$ components. Thus we have 
$\omega_x(x,y,z)=-\omega_x(x,-y,z)$, $\omega_y(x,y,z)=\omega_y(x,-y,z)$ and
$\omega_z(x,y,z)=-\omega_z(x,-y,z)$. Here $\omega_x, \; \omega_y , \; \omega_z$
are the $x$, $y$, and $z$ components of vorticity respectively. These symmetries 
allow us to compute only one quarter of the whole periodic cell.

A complete description of the initial condition is also given in \cite{HL06}.
There are a few misprints in the analytic expression of the initial 
condition given in \cite{Kerr93}. In our computations, we use the corrected 
version of Kerr's initial condition by comparing with Kerr's Fortran 
subroutine which was kindly provided to us by him.  A list of corrections
to these misprints is given in the Appendix of \cite{HL06}.

\begin{figure}
\begin{center}
\includegraphics[width=8cm]{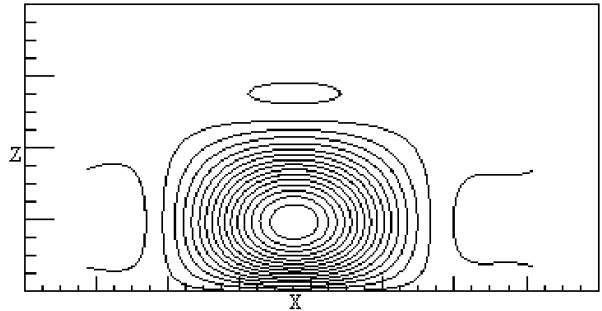}
\end{center}
\caption{The axial vorticity (the second component of vorticity) 
contours of the initial value on the symmetry plane. 
The vertical axis is the $z$-axis, and the horizontal axis is the $x$-axis.
\label{fig.vort-init}}
\end{figure}

\begin{figure}
\begin{center}
\includegraphics[width=8cm]{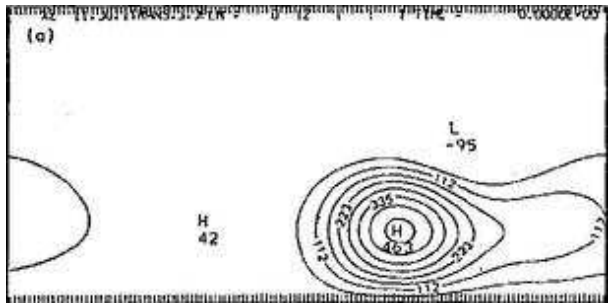}
\end{center}
\caption{Kerr's axial vorticity contours of the initial value on the symmetry plane.
The vertical axis is the $z$-axis, and the horizontal axis is the $x$-axis.
This is Fig. 2(a) of \cite{Kerr93}.
\label{fig.vort-init1}}
\end{figure}

\begin{figure}
\begin{center}
\includegraphics[width=8cm]{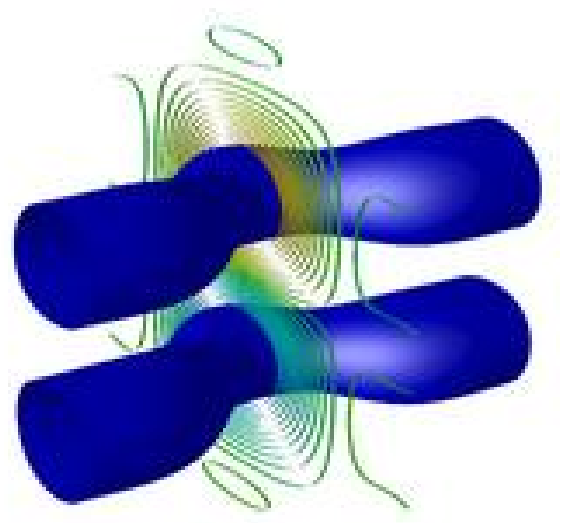}
\includegraphics[width=8cm]{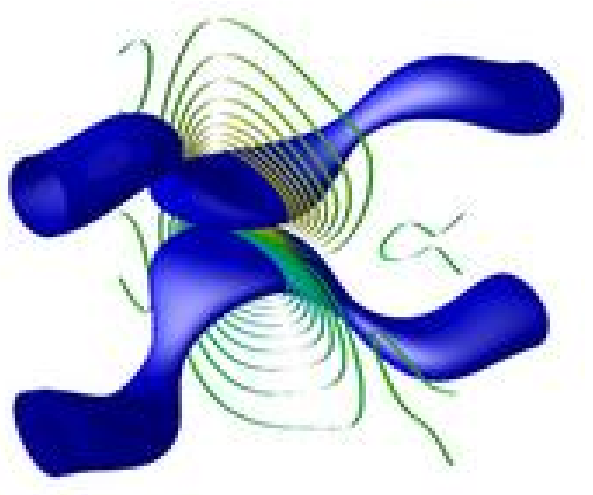}
\end{center}
\caption{The 3D view of the vortex tube for $t=0$ and $t=6$.
The tube is the isosurface at $60\%$ of the maximum vorticity.
The ribbons on the symmetry plane are the contours 
at other different values. 
\label{fig.vorttube}} 
\end{figure}

We should point out that due to the difference between Kerr's discretization 
strategies and ours in solving the 3D Euler equations, there is
some noticeable difference between the discrete initial condition 
generated by Kerr's discretization and the one generated 
by our pseudo-spectral discretization. In \cite{Kerr93}, Kerr
interpolated the initial condition from the uniform grid to the
Chebyshev grid along the $z$-direction and applied extra filtering. 
This interpolation and extra filtering, which were not provided
explicitly in \cite{Kerr93}, seem to introduce some numerical 
artifact to Kerr's discrete initial condition. According 
to \cite{Kerr93} (see the top paragraph of page 1729), ``An 
effect of the initial filter upon the vorticity contours at $t=0$ 
is a long tail in Fig. 2(a)''. Since our computations are
performed on a uniform grid using the pseudo-spectral approximations
in all three directions, we do not need to use any interpolation 
To demonstrate this slight difference between Kerr's discrete 
initial condition and ours, we plot the initial vorticity contours 
along the symmetry plane in Figure \ref{fig.vort-init} using our
spectral discretization in all three directions. As we can see, 
the initial vorticity contours in Figure \ref{fig.vort-init} are 
essentially symmetric. This is in contrast to the apparent asymmetry
in Kerr's initial vorticity contours as illustrated by
Figure \ref{fig.vort-init1}, which is Fig. 2(a) of \cite{Kerr93}. 
We also present the 3D plot of the vortex tubes at $t=0$ and $t=6$
respectively in Figure \ref{fig.vorttube}. We can see that 
the two initial vortex tubes are essentially symmetric. By time 
$t=6$, there is already a significant flattening near the center 
of the tubes. 

We exploit the symmetry properties of the solution in our computations,
and perform our computations on only a quarter of the whole domain.
Since the solution appears to be 
most singular in the $z$ direction, we allocate twice as many grid points 
along the $z$ direction than along the $x$ direction. The solution is 
least singular in the $y$ direction. We allocate the smallest resolution 
in the $y$ direction to reduce the computation cost. 
In our computations, 
two typical ratios in the resolution along the $x$, $y$ and $z$ directions 
are $3:2:6$ and $4:3:8$.
Our computations were carried out on the PC cluster LSSC-II in the
Institute of Computational Mathematics and Scientific/Engineering Computing
of Chinese Academy of Sciences and the Shenteng 6800 cluster in the Super
Computing Center of Chinese Academy of Sciences. The maximal memory
consumption in our computations is about 120 GBytes.

\section{Convergence study of the two numerical methods}

We use two numerical methods to compute the 3D Euler equations.
The first method is the pseudo-spectral method with the 2/3 
dealiasing rule. The second method is the pseudo-spectral method 
with a high order Fourier smoothing. The only difference between
these two methods is in the way we perform the cut-off of the
high frequency Fourier modes to control the aliasing error. 
If $\widehat{v}_k$ is the discrete Fourier transform of $v$, 
then we approximate the derivative of $v$ along the $x_j$ 
direction, $v_{x_j}$, by taking the discrete inverse Fourier 
transform of $i k_j \rho(2k_j/N_j) \widehat{v}_k$, where 
$k =(k_1,k_2,k_3)$ and $\rho$ is a high frequency Fourier 
cut-off function. Here $k_j$ is the wave number 
($|k_j| \leqslant N_j/2$) along the $x_j$ direction and 
$N_j$ is the total number of grid points along the $x_j$ 
direction. For the pseudo-spectral method with the 
2/3 dealiasing rule, the cut-off function $\rho$ is chosen
such that $\rho (x) = 1$ if $ |x| \leq 2/3$, and 
$\rho (x) = 0$ if $ 2/3 < |x| \leq 1$. For the pseudo-spectral
method with a high order smoothing, we choose the cut-off function
$\rho$ to be a smooth function of the form
$\rho(x) \equiv \exp(-\alpha |x|^m)$ with $\alpha=36$ and 
$m=36$. The time integration is performed 
using the classical fourth order Runge-Kutta method. Adaptive
time stepping is used to satisfy the CFL stability condition with 
CFL number equal to $\pi/4$. We use a sequence of
resolutions: $768\times 512\times 1536$,
$1024\times 768\times 2048$, and $1536\times 1024\times 3072$,
to demonstrate the convergence of our numerical computations.

\subsection{Comparison of the two methods}

\begin{figure}
\begin{center}
\includegraphics[width=8cm]{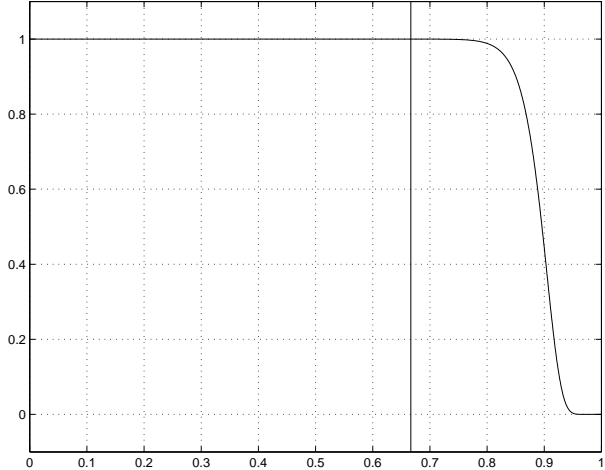}
\end{center}
\caption{The profile of the Fourier smoothing, $\exp(-36 (x)^{36})$,
as a function of $x$.
The vertical line corresponds to the cut-off mode using the $2/3$ 
dealiasing rule. We can see that using this Fourier smoothing we keep 
about $12 \sim 15\%$ more modes than those using the 2/3 dealiasing rule.
\label{fig.fourier_smoother}}
\end{figure}

It is interesting to make some comparison of the two spectral
methods we use. First of all, both methods are of spectral accuracy. 
The pseudo-spectral method with the 2/3 dealiasing rule has
been widely used in the computational fluid dynamics community.
It has the advantage of removing the aliasing error completely.
On the other hand, when the solution is nearly singular, the
Fourier spectrum typically decays very slowly. By cutting off
the last 1/3 of the high frequency modes along each direction 
abruptly, this can introduce oscillations in the physical solution 
due to the Gibbs phenomenon. In this paper, we will provide solid 
numerical evidences to demonstrate this effect. On the other
hand, the pseudo-spectral method with the high order Fourier 
smoothing is designed to keep the majority of the Fourier modes 
unchanged and remove the very high modes to avoid the aliasing error, 
see Fig. \ref{fig.fourier_smoother} for the profile of $\rho (x)$.
We choose $\alpha$ to be $36$ to guarantee that $\rho (2k_j/N_j)$ 
reaches the level of the round-off error ($O(10^{-16})$) at 
the highest modes. The order of smoothing, $m$, is chosen to be 
36 to optimize the accuracy of the spectral approximation, while 
still keeping the aliasing error under control. As we can see 
from Figure \ref{fig.fourier_smoother}, the effective modes in 
our computations are about $12 \sim 15\%$ more than those using 
the standard $2/3$ dealiasing rule. Retaining part of the effective
high frequency Fourier modes beyond the traditional 2/3 cut-off 
position is a special feature of the second method. 

\begin{figure}
\begin{center}
\includegraphics[width=12cm,height=6cm]{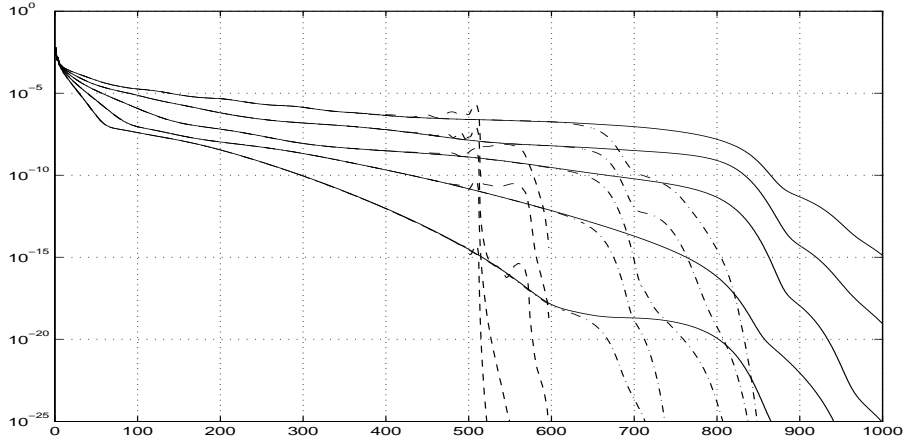}
\end{center}
\caption{The enstrophy spectra versus wave numbers. We compare the enstrophy 
spectra obtained using the high order 
Fourier smoothing method with those using the 2/3 dealiasing rule. The dashed 
lines and dashed-dotted lines are the enstrophy spectra with the
resolution $768\times 512\times 1536$ using the 2/3 dealiasing rule and the
 Fourier smoothing, respectively. The solid lines are the enstrophy spectra
with resolution $1024\times 768\times 2048$ obtained using the high order
Fourier smoothing. The times for the spectra lines 
are at $t= 15, 16, 17, 18, 19$ respectively. 
\label{fig.enstrophy-spec-comp}}
\end{figure}

To compare the performance of the two methods, we perform 
a careful convergence study for the two methods. In 
Figure \ref{fig.enstrophy-spec-comp}, we compare the Fourier 
spectra of the enstrophy obtained by using the pseudo-spectral 
method with the $2/3$ dealiasing rule with those obtained by 
the pseudo-spectral method with the high order smoothing. For
a fixed resolution $768\times 512\times 1536$, we can see
that the Fourier spectra obtained by the pseudo-spectral
method with the high order smoothing retains more effective
Fourier modes than those obtained by the spectral method with 
the $2/3$ dealiasing rule. This can be seen by comparing the 
results with the corresponding computations using a higher 
resolution $1024 \times 768 \times 2048$. Moreover, the 
pseudo-spectral method with the high order Fourier smoothing 
does not give the spurious oscillations in the Fourier spectra 
which are present in the computations using the $2/3$ dealiasing 
rule near the $2/3$ cut-off point. 

\begin{figure}
\begin{center}
\includegraphics[width=12cm,height=6cm]{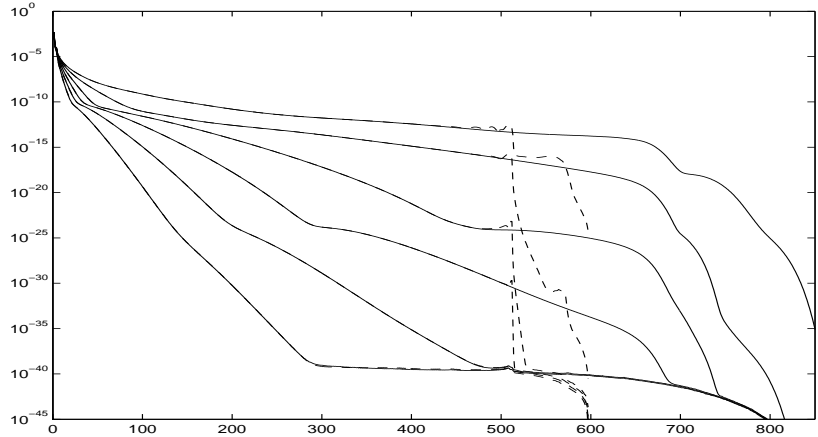}
\end{center}
\caption{The energy spectra versus wave numbers. We compare the energy
spectra obtained using the high order
Fourier smoothing method with those using the 2/3 dealiasing rule. The dashed
lines and solid lines are the energy spectra with the
resolution $768\times 512\times 1536$ using the 2/3 dealiasing rule and the
 Fourier smoothing, respectively. The times for the spectra lines
are at $t= 8, 10, 12, 14, 16, 18$ respectively.
\label{fig.energy-spec-comp}}
\end{figure}

We perform further comparison of the two methods using the
same resolution. In  Figure \ref{fig.energy-spec-comp},
we plot the energy spectra computed by the two methods 
using resolution $768\times 512\times 1536$. We can see that
there is almost no difference in the Fourier spectra 
generated by the two methods in early times, $t=8, 10$,
when the solution is still relatively smooth. The difference 
begins to show near the cut-off point when the Fourier spectra 
raise above the round-off error level starting from $t=12$. 
We can see that the spectra computed by the pseudo-spectral
method with the 2/3 dealiasing rule introduces noticeable
oscillations near the 2/3 cut-off point. The spectra
computed by the pseudo-spectral method with the high
order smoothing, on the other hand, extend smoothly
beyond the 2/3 cut-off point. As we see from 
Figure \ref{fig.enstrophy-spec-comp}, a significant portion 
of those Fourier modes beyond the 2/3 cut-off position 
are still accurate. In the next subsection, we will 
demonstrate by a careful resolution study that the 
pseudo-spectral method with the high order smoothing
indeed offers better accuracy than the pseudo-spectral 
method with the 2/3 dealiasing rule.

\begin{figure}
\begin{center}
\includegraphics[width=10cm,height=6cm]{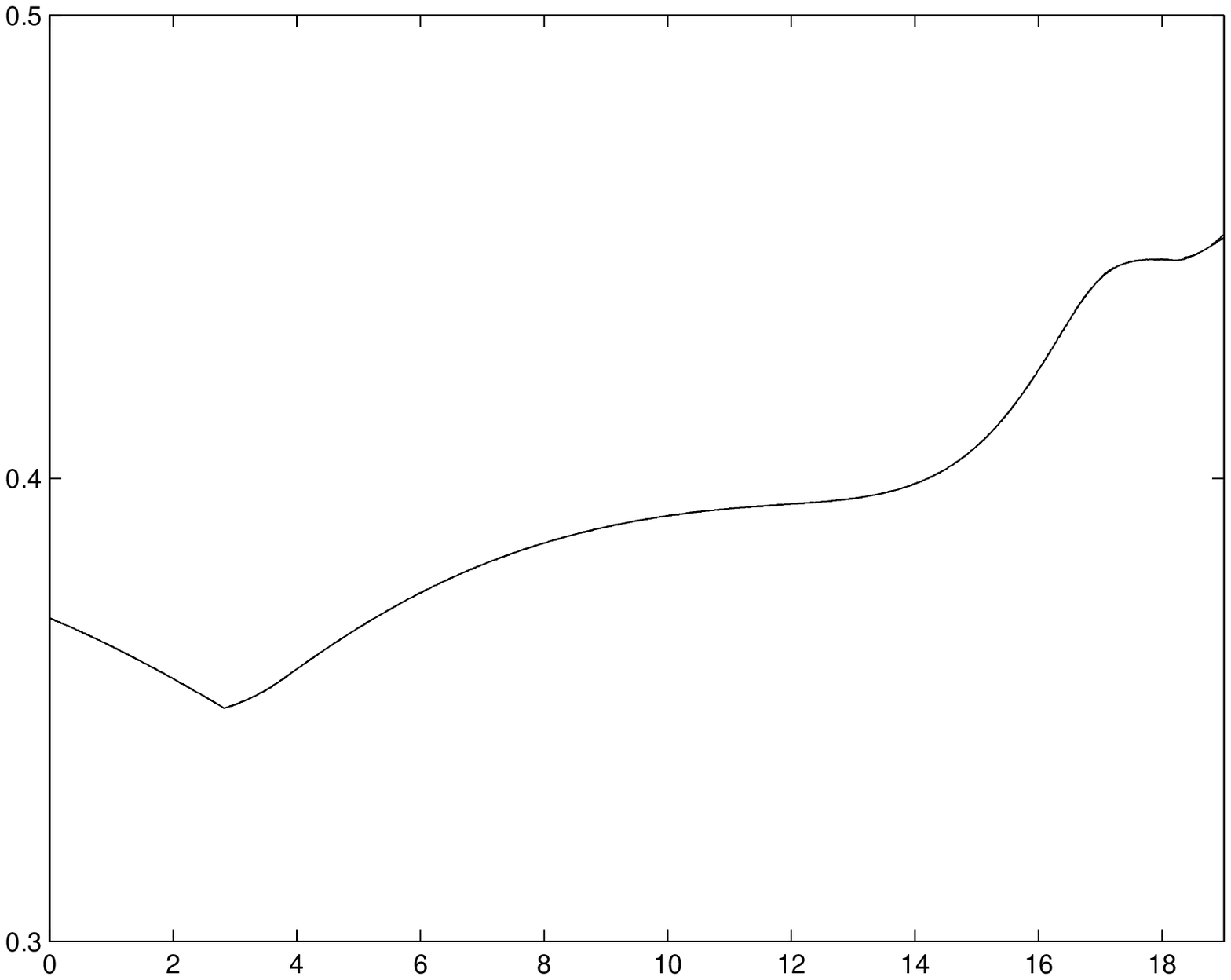}
\end{center}
\caption{Comparison of maximum velocity as a function of time
computed by two methods. The solid line represents the solution
obtained by the pseudo-spectral method with the high order smoothing,
and the dashed line represents the solution obtained by
the pseudo-spectral method with the 2/3 dealiasing rule.
The resolution is $768\times 512\times 1536$ for both methods.
\label{fig.max-u-comp-768}}
\end{figure}

\begin{figure}
\begin{center}
\includegraphics[width=10cm]{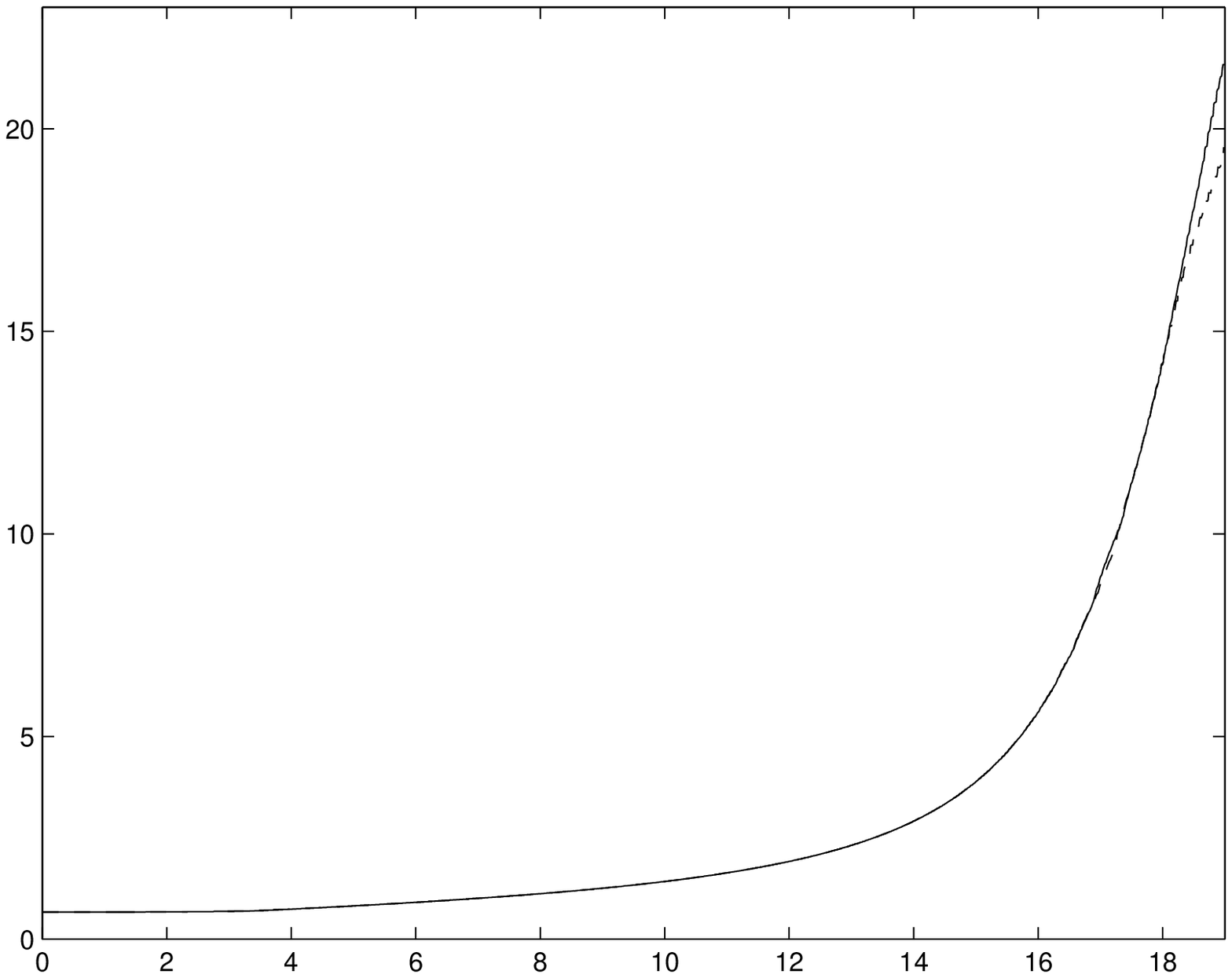}
\end{center}
\caption{Comparison of maximum vorticity as a function of time
computed by two methods. The solid line represents the solution
obtained by the pseudo-spectral method with the high order smoothing,
and the dashed line represents the solution obtained by
the pseudo-spectral method with the 2/3 dealiasing rule.
The resolution is $768\times 512\times 1536$ for both methods.
\label{fig.max-vort-comp-768}}
\end{figure}

\begin{figure}
\begin{center}
\includegraphics[width=7cm, angle=270]{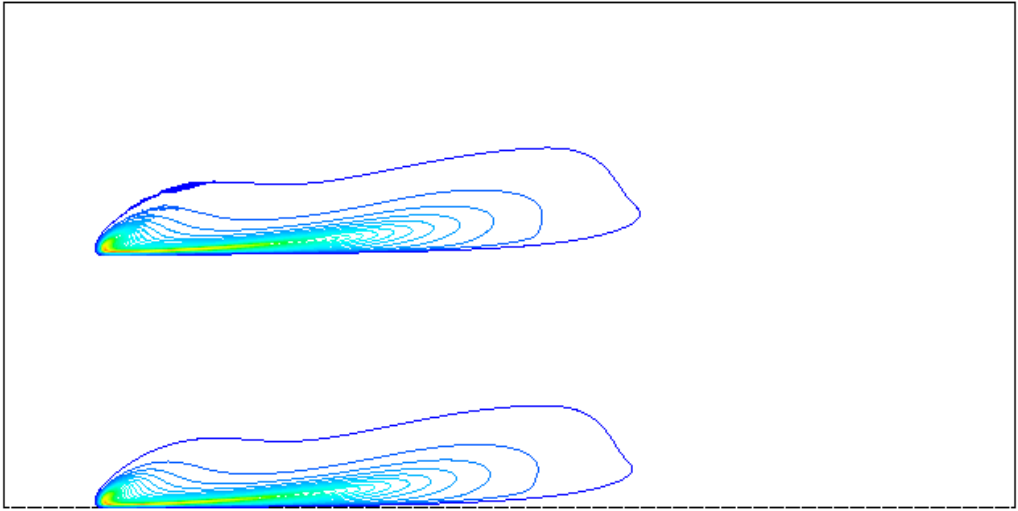}
\end{center}
\caption{Comparison of axial vorticity contours at $t=17$ 
computed by two methods. The picture on the top is the solution
obtained by the pseudo-spectral method with the 2/3 dealiasing rule,
which is shifted by a distance of $\pi$ in $z$ direction,
and the picture on the bottom is the solution obtained by
the pseudo-spectral method with the high order smoothing.
The resolution is $768\times 512\times 1536$ for both methods. The box
is the whole $x-z$ computational domain $[-2\pi, 2 \pi]\times [0,
2\pi]$. 
\label{fig.vort-cont-comp-768-17}}
\end{figure}

\begin{figure}
\begin{center}
\includegraphics[width=6cm,angle=270]{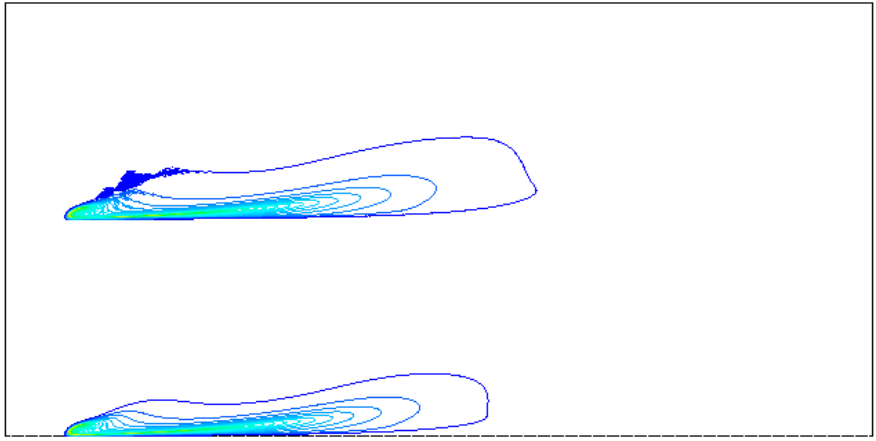}
\end{center}
\caption{Comparison of axial vorticity contours at $t=18$
computed by two methods. This figure has the same layout as Figure 
\ref{fig.vort-cont-comp-768-17}. The top picture uses
the 2/3 dealiasing rule, while the bottom picture uses
the high order smoothing. The resolution 
is $768\times 512\times 1536$ for both methods. 
\label{fig.vort-cont-comp-768-18}}
\end{figure}

Similar comparison can be made in the physical space for the 
velocity field and the vorticity. In Figure \ref{fig.max-u-comp-768}, 
we compare the maximum velocity as a function of time computed by 
the two methods using resolution $768\times 512\times 1536$. The 
two solutions are almost indistinguishable. In Figure 
\ref{fig.max-vort-comp-768}, we plot the maximum vorticity
as a function of time. The two solutions agree very well up to 
$t=18$. The solution obtained by the pseudo-spectral method with 
the 2/3 dealiasing rule grows slower from $t=18$ to $t=19$. 
To understand why the two solutions start to deviate from
each other toward the end, we examine the contour plot of the 
axial vorticity in Figures \ref{fig.vort-cont-comp-768-17}
and \ref{fig.vort-cont-comp-768-18}. As we can see,
the vorticity computed by the pseudo-spectral method
with the 2/3 dealiasing rule already develops small
oscillations at $t=17$. The oscillations grow bigger 
by $t=18$. We note that the oscillations in the axial
vorticity contours concentrate near the region where the 
magnitude of vorticity is close to zero. On the other hand, 
the solution computed by the spectral method with the 
high order smoothing is still quite smooth. 

\subsection{Resolution study for the two methods}

In this subsection, we perform a resolution study for the
two numerical methods using a sequence of resolutions.
For the pseudo-spectral method with the high order smoothing,
we use the resolutions $768\times 512\times 1536$, 
$1024\times 768\times 2048$, and $1536\times 1024\times 3072$ 
respectively. Except for the computation on the largest resolution 
$1536\times 1024\times 3072$, all computations are carried out from 
$t=0$ to $t=19$. The computation on the final resolution 
$1536\times 1024\times 3072$ is started from $t=10$ with the 
initial condition given by the computation with the resolution
$1024\times 768\times 2048$. For the pseudo-spectral method
with the 2/3 dealiasing rule, we use 
the resolutions $512 \times 384 \times 1024$, 
$768\times 512\times 1536$ and $1024\times 1024\times 2048$
respectively. The computations using the first two resolutions
are carried out from $t=0$ to $t=19$ while the computation
on the largest resolution $1024\times 1024\times 2048$
is started at $t=15$ with the initial condition given by
the computation with resolution $512\times 512\times 1024$.

\begin{figure} \begin{center}
\includegraphics[width=12cm,height=6cm]{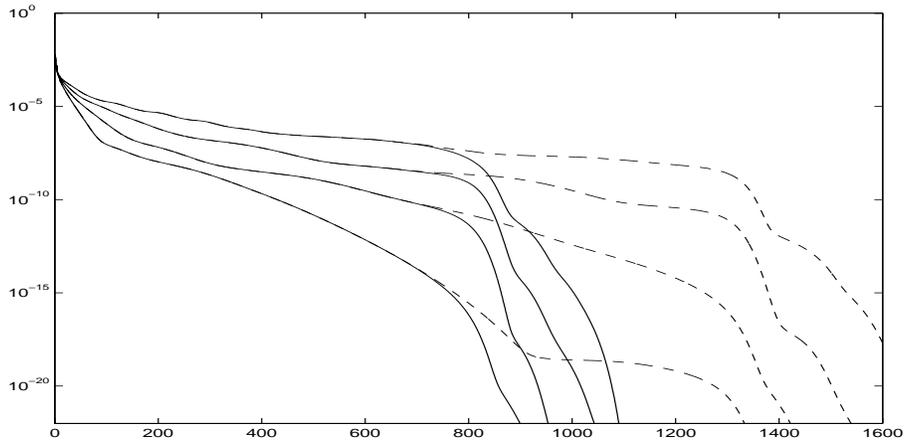}
\end{center}
\caption{Convergence study for enstrophy spectra obtained by
the pseudo-spectral method with high order smoothing using 
different resolutions.
The dashed lines and the solid lines are the enstrophy spectra on
resolution $1536\times 1024\times 3072$ and $1024\times 768\times
2048$, respectively. The times for the lines from bottom to top are
$t=16, 17, 18, 19$.
\label{fig.enstrophy_spec_1}}
\end{figure}

\begin{figure}
\begin{center}
\includegraphics[width=12cm,height=6cm]{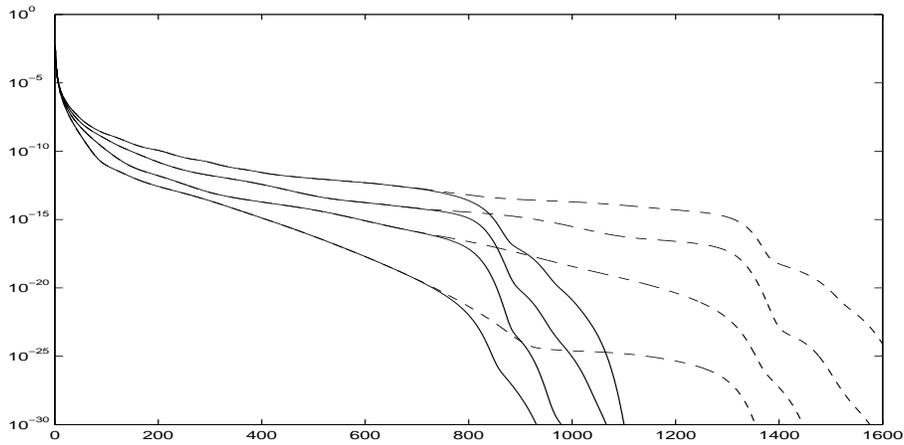}
\end{center}
\caption{Convergence study for energy spectra obtained by
the pseudo-spectral method with high order smoothing using different 
resolutions.
The dashed lines and the solid lines are the energy spectra on resolution
$1536\times 1024\times 3072$ and $1024\times 768\times 2048$,
respectively. The times for the lines from bottom to top are $t=16, 17,
18, 19$.
\label{fig.energy_spec_1}}
\end{figure}

First, we perform a convergence study of the enstrophy and 
energy spectra for the pseudo-spectral method with the high
order smoothing at later times (from $t=16$ to $t=19$) using two 
largest resolutions $1024\times 768\times 2048$, and 
$1536\times 1024\times 3072$. The results are given in Figures 
\ref{fig.enstrophy_spec_1} and \ref{fig.energy_spec_1} respectively. 
They clearly demonstrate the spectral convergence of the spectral 
method with the high order smoothing.

\begin{figure}
\begin{center}
\includegraphics[width=9cm]{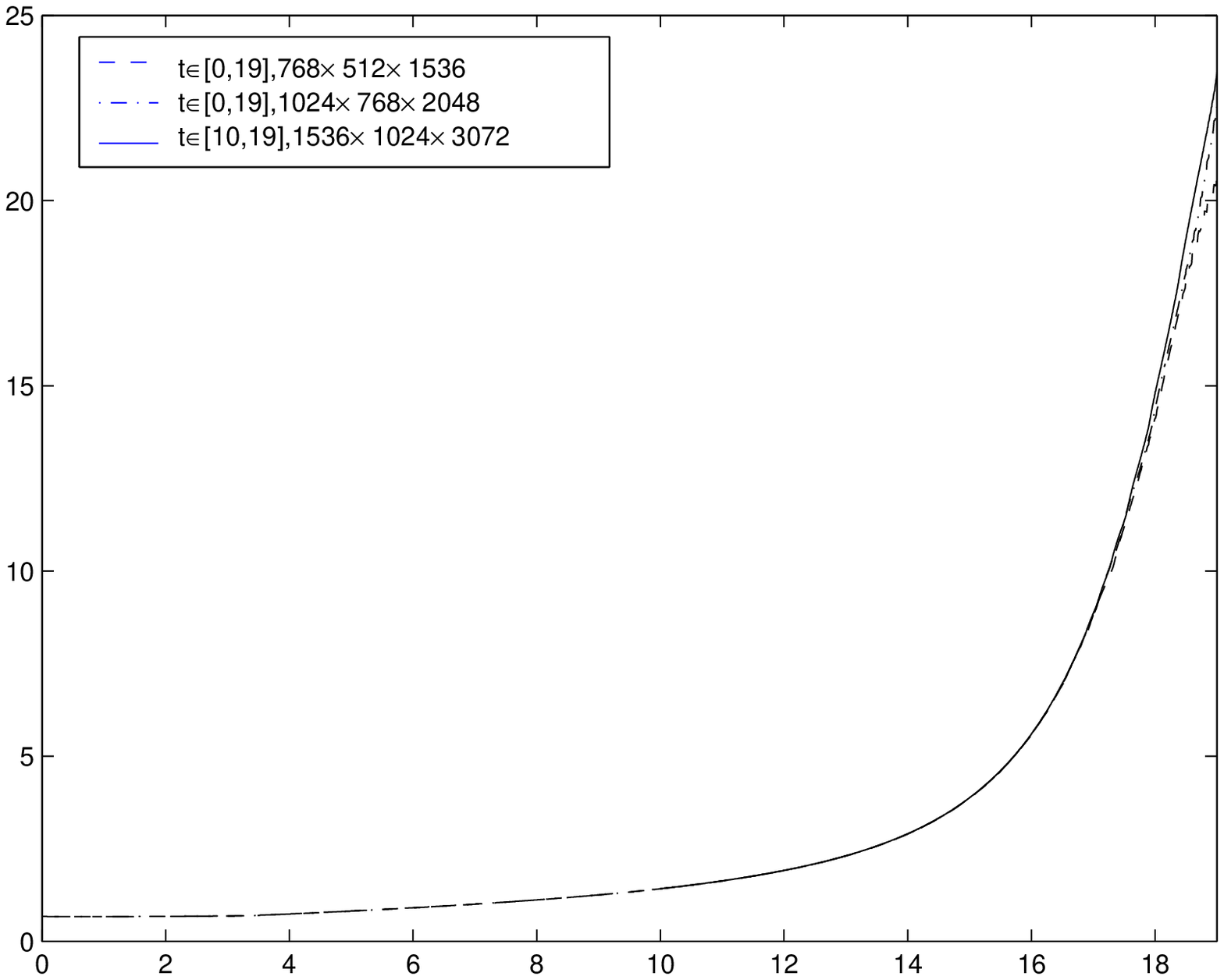}
\end{center} 
\caption{The maximum vorticity $\|\vec{\omega}\|_\infty$ in time 
computed by the pseudo-spectral method with high order 
smoothing using
different resolutions.
\label{fig.omega}}
\end{figure}

To further demonstrate the accuracy of our computations, we compare the 
maximum vorticity obtained by the pseudo-spectral method with the 
high order smoothing for three different resolutions:
$768\times 512\times 1536$, $1024\times 768\times 2048$, and
$1536\times 1024\times 3072$ respectively.
The result is plotted in Figure \ref{fig.omega}. Two conclusions
can be made from this resolution study. First, by comparing
Figure \ref{fig.omega} with Figure \ref{fig.max-vort-comp-768},
we can see that the pseudo-spectral method with the high order
smoothing is indeed more accurate than the pseudo-spectral method
with the 2/3 dealiasing rule for a given resolution. Secondly, the 
resolution $768\times 512\times 1536$ is not good enough to resolve 
the nearly singular solution at later times. However, we observe
that the difference of the numerical solution obtained by
the resolution $1024\times 768\times 2048$ is very close to
that obtained by the resolution $1536\times 1024\times 3072$.
This indicates that the vorticity is reasonably well-resolved
by our largest resolution $1536\times 1024\times 3072$. 

\begin{figure} 
\begin{center}
\includegraphics[width=9cm]{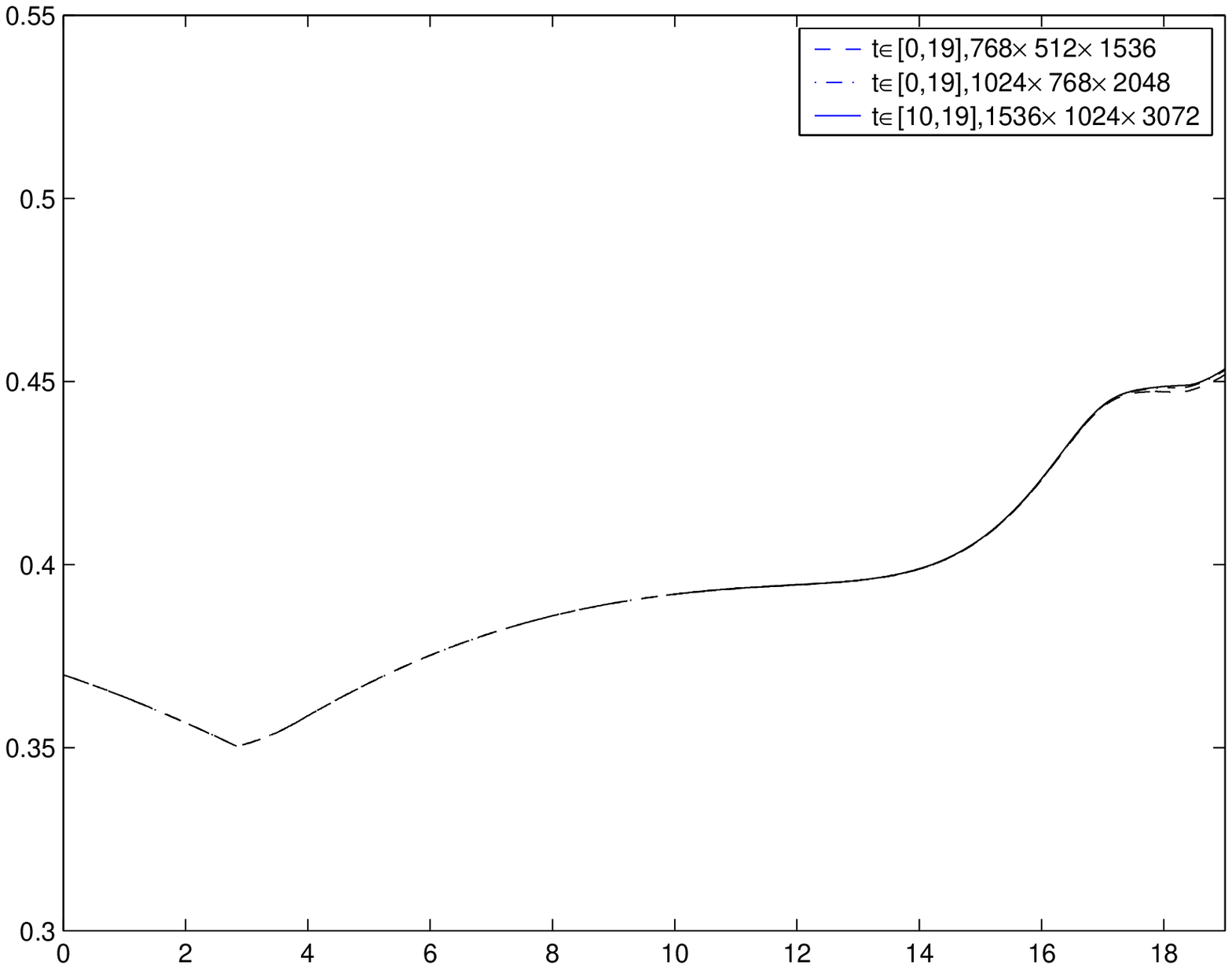}
\end{center}
\caption{Maximum velocity $\|\vec{u}\|_\infty$ in time
computed by the pseudo-spectral method with high order
smoothing using different resolutions.
\label{fig.velocity}} 
\end{figure}

We have also performed a similar resolution study for the maximum
velocity in Figure \ref{fig.velocity}. The solutions obtained by
the two largest resolutions are almost indistinguishable,
which suggests that the velocity is well-resolved by
our largest resolution $1536\times 1024\times 3072$.

\begin{figure} 
\begin{center}
\includegraphics[width=12cm,height=6cm]{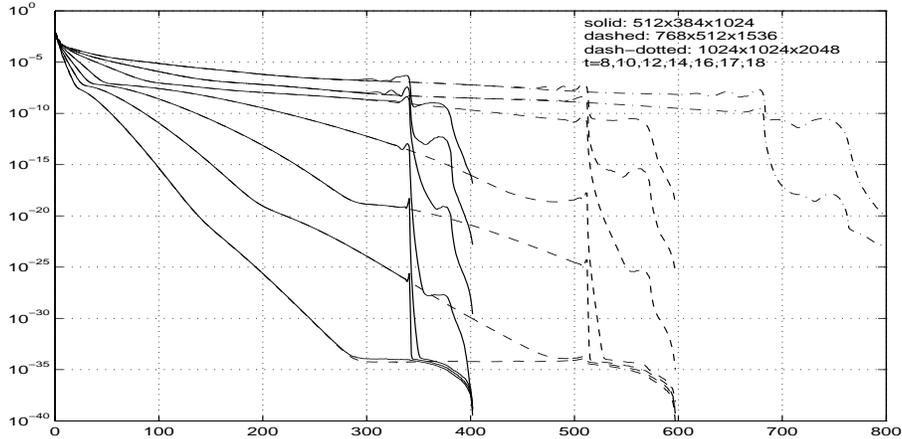}
\end{center}
\caption{Convergence study for enstrophy spectra obtained by
the pseudo-spectral method with the 2/3 dealiasing rule
using different resolutions. The solid line is computed with
resolution $512\times 384\times 1024$, the dashed line is 
computed with resolution $786\times 512\times 1536$, and
the dashed-dotted line is computed with resolution 
$1024\times 1024\times 2048$. The times for the lines 
from bottom to top are $t=8, 10, 12, 14, 16, 17, 18$.
\label{fig.enstrophy_spec_2}}
\end{figure}

Next, we perform a similar resolution study for the pseudo-spectral
method with the 2/3 dealiasing rule. The results are very similar
to the ones we have obtained for the pseudo-spectral method 
with the high order smoothing. Here we just present a few
representative results. In Figure \ref{fig.enstrophy_spec_2},
we plot the enstrophy spectra for a sequence of times from
$t=8$ to $t=18$ using different resolutions. The resolutions 
we use here are 
$512\times 384\times 1024$, $786\times 512\times 1536$,
and $1024\times 1024\times 2048$ respectively. If we 
compare the Fourier spectra at $t=17$ and $t=18$ (the 
last two curves in Figure \ref{fig.enstrophy_spec_2}), we 
clearly observe convergence of the enstrophy spectra as
we increase our resolutions. On the other hand, the 
decay of the enstrophy spectra becomes very slow at 
later times. The oscillations near the 2/3 cut-off point
become more and more pronounced as time increases. 
This abrupt cut-off of high frequency spectra introduces 
some oscillations in the vorticity contours at later times.

\begin{figure}
\begin{center}
\includegraphics[width=12cm,height=6cm]{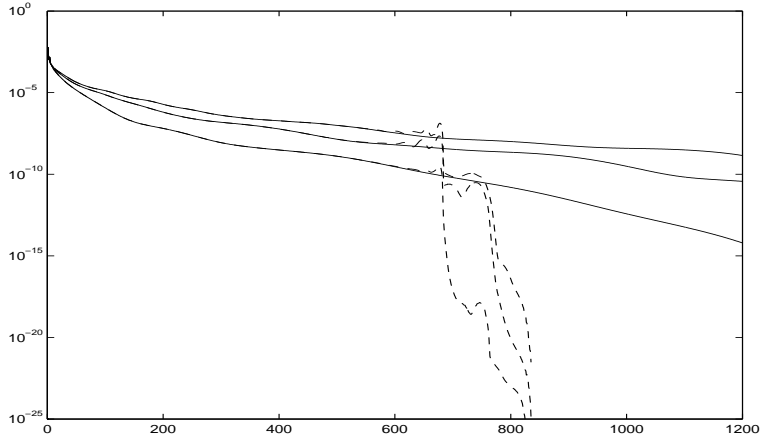}
\end{center}
\caption{The enstrophy spectra versus wave numbers. We compare the enstrophy
spectra obtained using the high order
Fourier smoothing method with those using the 2/3 dealiasing rule. The dashed
lines lines are the enstrophy spectra using the 2/3 dealiasing rule with resolution 
$1024\times 1024\times 2048$, and the solid lines are the spectra with
resolution $1536\times 1024\times 3072$ using the Fourier smoothing. 
The times for the spectra lines are at $t= 17, 18, 18.5$ respectively.
\label{fig.enstrophy-spec-comp-1024}}
\end{figure}

To demonstrate that the two numerical methods converge
to the same solution when the solution is nearly singular, 
we compare the enstrophy spectra computed by the two numerical 
methods at later times using the largest resolutions
that we can afford. For the pseudo-spectral method
with the high order smoothing, we use resolution
$1536\times 1024\times 3072$. For the pseudo-spectral
method with the 2/3 dealiasing rule, we use resolution
$1024\times 1024\times 2048$. 
In Figure \ref{fig.enstrophy-spec-comp-1024}, we plot 
the enstrophy spectra for $t=17$, 18, and 18.5 respectively.
We observe that the two methods give excellent agreement for 
those Fourier modes that are not affected by the high frequency 
cut-off. This shows that the two numerical methods converge 
to the same solution with spectral accuracy. 

\begin{figure}
\begin{center}
\includegraphics[width=9cm]{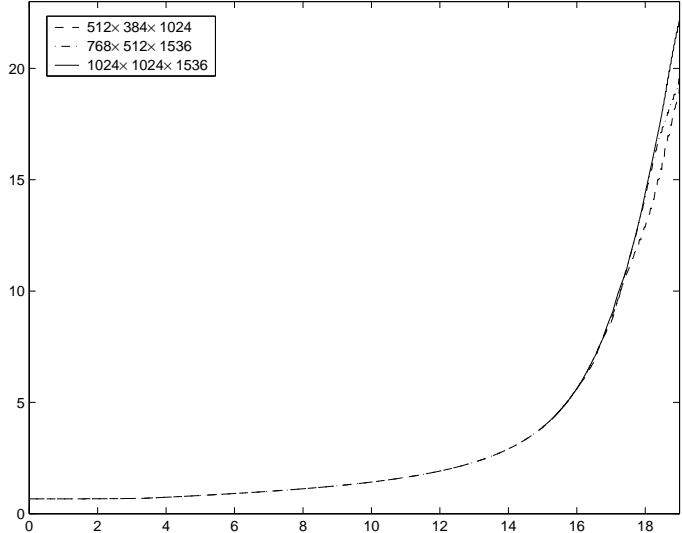}
\end{center}
\caption{The maximum vorticity $\|\vec{\omega}\|_\infty$ in time
computed by the pseudo-spectral method with the 2/3 dealiasing
rule using different resolutions.
\label{fig.omega_2}} 
\end{figure}

We have performed a similar convergence study for the
pseudo-spectral method with the 2/3 dealiasing in the 
physical space for the maximum vorticity. The result is
given in Figure \ref{fig.omega_2}. As we can see, 
the computation with a higher resolution gives faster 
growth in the maximum vorticity. This is also what 
we observed earlier for the pseudo-spectral method 
with the high order smoothing. As we will see in
the next section, the maximum vorticity grows 
almost like doubly exponential in time. To capture this
rapid dynamic growth of maximum vorticity, we must
have sufficient resolution to resolve the nearly 
singular solution of the Euler equations at later 
times. 

The resolution study given by Figure \ref{fig.omega_2}
suggests that the maximum vorticity is reasonably 
resolved by resolution $768\times 512\times 1536$ before 
$t=18$. It is interesting to note that at $t=17$, small
oscillations have already appeared in the vorticity
contours in the region where the magnitude of vorticity
is small, see Figure \ref{fig.vort-cont-comp-768-17}.
Apparently, the small oscillations in the region
where the vorticity is close to zero in magnitude have 
not yet polluted the accuracy of the maximum vorticity 
in a significant way. Note that there is no oscillation
developed in the vorticity contours obtained by the 
pseudo-spectral method with the high order smoothing
at this time. From Figure \ref{fig.max-vort-comp-768}, 
we know that the maximum vorticity computed by the two
methods agrees reasonably well with each other before $t=18$. 
This shows that the two methods can still approximate
the maximum vorticity reasonably well with resolution 
$768\times 512\times 1536$ before $t=18$.

The resolution study given by Figure \ref{fig.omega_2} 
also suggests that the the computation obtained by the 
pseudo-spectral method with the 2/3 dealiasing rule using 
resolution $768\times 512\times 1536$ is significantly
under-resolved after $t=18$. This is also supported by 
the appearance of the relatively large oscillations in 
the vorticity contours at $t=18$ from 
Figure \ref{fig.vort-cont-comp-768-18}. It is interesting 
to note from Figure \ref{fig.max-vort-comp-768} that 
the computational results obtained by the two methods 
with resolution $768\times 512\times 1536$
begin to deviate from each other precisely around $t=18$. By 
comparing the result from Figure \ref{fig.max-vort-comp-768} 
with that from Figure \ref{fig.omega_2}, we confirm again 
that for a given resolution, the pseudo-spectral method with 
the high order smoothing gives a more accurate approximation
than the pseudo-spectral method with the 2/3 dealiasing rule.

                                                                                         
\section{Analysis of computational results}

In this section, we will present a series of numerical results 
to reveal the nature of the nearly singular solution of the 
3D Euler equations, and compare our results with those obtained 
by Kerr in \cite{Kerr93,Kerr05}. Based on the convergence study we have 
performed in the previous section, we will present only those 
numerical results which are computed by the pseudo-spectral 
method with the high order smoothing using the largest resolution 
$1536\times 1024\times 3072$.

\subsection{Review of Kerr's results}

In \cite{Kerr93}, Kerr presented numerical evidence which suggested
a finite time singularity of the 3D Euler equations for two perturbed 
antiparallel vortex tubes. He used a pseudo-spectral discretization 
in the $x$ and $y$ directions, and a Chebyshev method in the $z$ 
direction with resolution of order $512\times 256 \times 192$. His 
computations showed that the growth of the peak vorticity, the 
peak axial strain, and the enstrophy production obey 
$(T-t)^{-1}$ with $T = 18.9$. Kerr stated in his paper 
\cite{Kerr93} (see page 1727) that his numerical results shown 
after $t=17$ and up to $t=18$ were ``not part of the primary evidence 
for a singularity'' due to the lack of sufficient numerical 
resolution and the presence of noise in the numerical solutions. 
In his recent paper \cite{Kerr05} (see also \cite{Kerr97,Kerr99}), 
Kerr applied a high wave number filter to the data obtained in his 
original computations to ``remove the noise that masked the structures 
in earlier graphics'' presented in \cite{Kerr93}. With this filtered 
solution, he presented some scaling analysis of the numerical solutions 
up to $t=17.5$. The velocity field was shown to blow up like 
$O((T-t)^{-1/2})$ with $T$ being revised to $T=18.7$. 

\subsection{Maximum vorticity growth}

From the resolution study we present in Figure \ref{fig.omega},
we find that the maximum vorticity increases rapidly from the 
initial value of $0.669$ to $23.46$ at the final time $t=19$, 
a factor of 35 increase from its initial value. Kerr's 
computations predicted a finite time singularity at $T=18.7$. 
Our computations show no sign of finite time blowup of the 3D Euler 
equations up to $T=19$, beyond the singularity time predicted by 
Kerr. We use three different resolutions, i.e. 
$768\times 512\times 1536$, $1024 \times 768 \times 2048$,
and $1536 \times 1024 \times 3072$ respectively in our 
computations. As we can see, the agreement between the two 
successive resolutions is very good with only mild disagreement 
toward the end of the computations. This indicates that a 
very high space resolution is indeed needed to capture the rapid 
growth of maximum vorticity at the later stage of the computations. 

\begin{figure}
\begin{center}
\includegraphics[width=8cm]{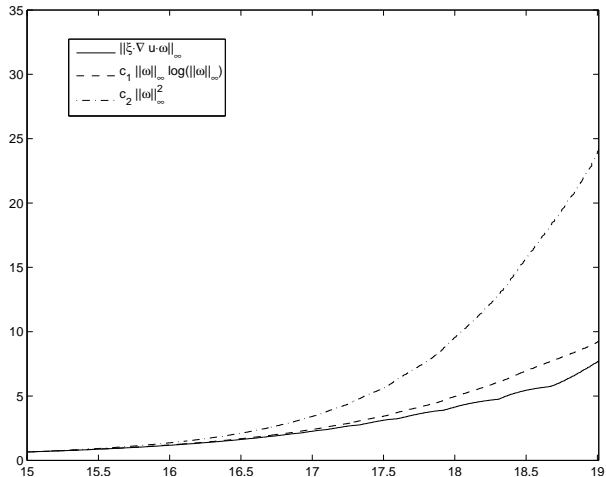}
\end{center}
\caption{Study of the vortex stretching term in time, resolution $1536\times 
1024\times 3072$. We take $c_1 = 1/8.128$, $c_2 = 1/23.24$ to match the 
same starting value for all three plots. 
\label{fig.growth_rate}}
\end{figure}

\begin{figure}
\begin{center}
\includegraphics[width=8cm]{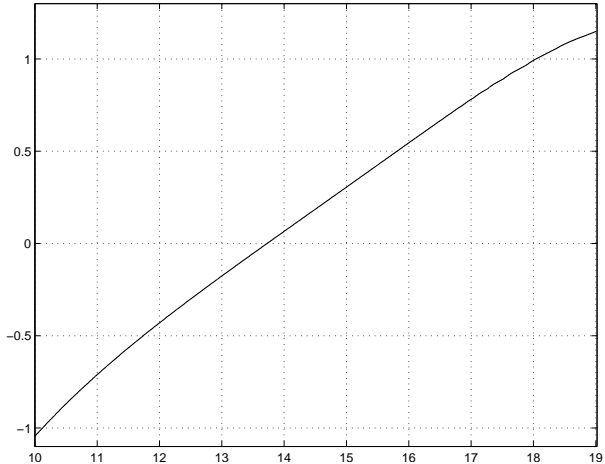}
\end{center}
\caption{The plot of $ \log \log \|\omega\|_\infty$ vs time, resolution $1536\times 1024
\times 3072$. 
\label{fig.omega_loglog}}
\end{figure}

In order to understand the nature of the dynamic growth in
vorticity, we examine the degree of nonlinearity in the vortex 
stretching term. In Figure \ref{fig.growth_rate}, we plot
the quantity, 
$\|\xi \cdot \nabla \vec{u} \cdot \vec{\omega}\|_\infty$,
as a function of time, where $\xi$ is the unit vorticity vector.
If the maximum vorticity indeed blew up like $O((T-t)^{-1})$, 
as alleged in \cite{Kerr93}, this quantity should have been
quadratic as a function of maximum vorticity. We find that 
there is tremendous cancellation in this vortex stretching 
term. It actually grows slower than
$C\|\vec{\omega} \|_\infty \log (\|\vec{\omega} \|_\infty )$,
see Figure \ref{fig.growth_rate}.
It is easy to show that such weak nonlinearity in vortex stretching
would imply only doubly exponential growth in the maximum vorticity.
Indeed, as demonstrated by Figure \ref{fig.omega_loglog}, the
maximum vorticity does not grow faster than doubly exponential
in time. In fact, the growth slows down toward the end of the
computation, which indicates that there is stronger cancellation 
taking place in the vortex stretching term.

\begin{figure}
\begin{center}
\includegraphics[width=8cm]{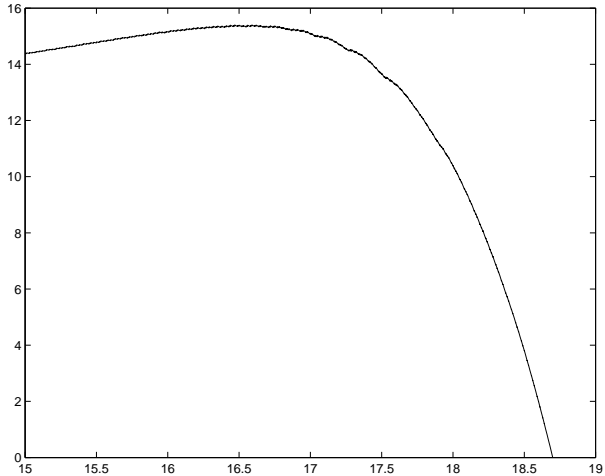}
\end{center}
\caption{Scaling constant in time for the fitting $\|\omega\|_\infty \approx c/(T-t)$,
$T=18.7$.
\label{fig.scaling_const}}
\end{figure}

We remark that for vorticity that grows as rapidly as doubly exponential in time,
one may be tempted to fit the maximum vorticity growth by $c/(T-t)$ for some 
$T$. Indeed, if we choose $T=18.7$ as suggested by Kerr in \cite{Kerr05}, we
find a reasonably good fit for the maximum vorticity as a function of
$c/(T-t)$ for the period $15 \le t \le 17$. We plot the scaling constant
$c$ in Figure \ref{fig.scaling_const}. As we can see, $c$ is close to a 
constant for $15 \le t \le 17$. To conclude that the 3D Euler equations
indeed develop a finite time singularity, one must demonstrate that such 
scaling persists as $t$ approaches to $T$. As we can see from Figure 
\ref{fig.scaling_const}, the scaling constant $c$ decreases rapidly to
zero as $t$ approaches to the alleged singularity time $T$. Therefore,
the fitting of $\| \vec{\omega}\|_\infty \approx O((T-t)^{-1})$ is not correct
asymptotically.

\subsection{Velocity profile}

One of the important findings of our computations is that the velocity 
field is actually bounded by 1/2 up to $T=19$. This is in contrast to Kerr's 
computations in which the maximum velocity was shown to blow up like 
$O((T-t)^{-1/2})$ \cite{Kerr97,Kerr05}. We plot 
the maximum velocity as a function of time using different 
resolutions in Figure \ref{fig.velocity}. The computation 
obtained by resolution $1024\times 768 \times 2048$ and 
the one obtained by resolution $1536\times 1024\times 3072$ are
almost indistinguishable. The fact that the velocity field is 
bounded is significant. With the velocity field being bounded, 
the non-blowup theory of Deng-Hou-Yu \cite{DHY05a} can be applied, 
which implies non-blowup of the 3D Euler equations up to $T$.
We refer to \cite{HL06} for more discussions.

\subsection{Local vorticity structure}

\begin{figure}
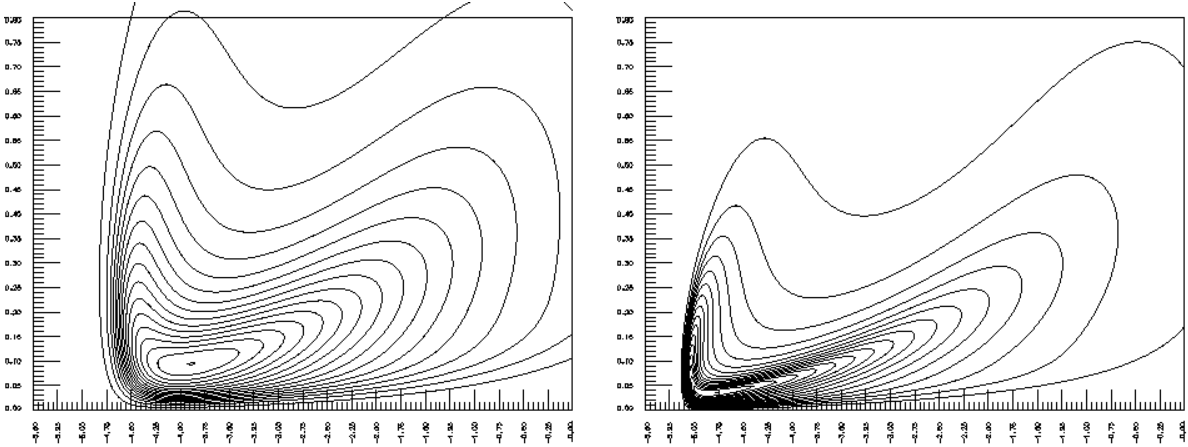

\begin{center}
\includegraphics[width=8cm]{pics/scaled_contour_t=15.epsf}
\includegraphics[width=8cm]{pics/scaled_contour_t=17.epsf}
\end{center}
\caption{The contour of axial vorticity around the maximum vorticity on the symmetry 
plane at $t=15$ (on the left) and $t=17$ (on the right). 
The vertical axis is the $z$-axis, and the horizontal axis is the $x$-axis.
The figure is scaled in $z$ direction by a factor of $5$ to compare with 
Figure 4 in \cite{Kerr93}. 
\label{fig.scaled_local_struc}}
\end{figure}

\begin{figure}
\begin{center}
\includegraphics[width=8cm]{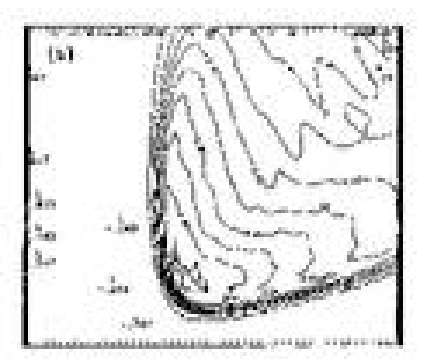}
\includegraphics[width=8cm]{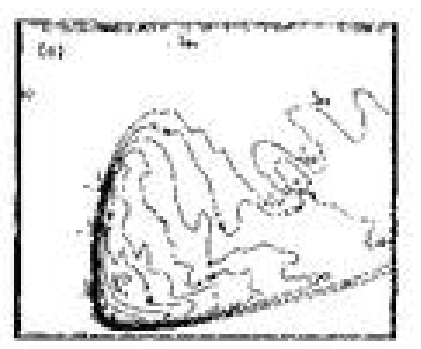}
\end{center}
\caption{Kerr's axial vorticity contours on the 
symmetry plane at $t=15$ (on the left) and $t=17$ (on the right). 
These are from Figure 4 in \cite{Kerr93}. 
\label{fig.kerr-scaled_local_struc}}
\end{figure}

In this subsection, we would like to examine the local vorticity structure
near the region of the maximum vorticity. To illustrate the development in
the symmetry plane, we show a series of vorticity contours near the region
of the maximum vorticity at late times in a manner similar to the results
presented in \cite{Kerr93}. For some reason, Kerr scaled his
axial vorticity contours by a factor of 5 along the $z$-direction.
Noticeable oscillations already develop in Kerr's axial vorticity
contours at $t=15$ and $t=17$, see Figure \ref{fig.kerr-scaled_local_struc}.
To compare with Kerr's results, we scale the vorticity contours in 
the $x-z$ plane by a factor of 5 in the $z$-direction. The results 
at $t=15$ and $t=17$ are plotted in Figure \ref{fig.scaled_local_struc}. 
The results are in qualitative agreement with Kerr's results, except 
that our computations are better resolved and do not suffer from the 
noise and oscillations which are present in Kerr's vorticity contours. 

\begin{figure}
\begin{center}
\includegraphics[width=8cm]{pics/fig4.noscale.t=17.5.epsf}
\includegraphics[width=8cm]{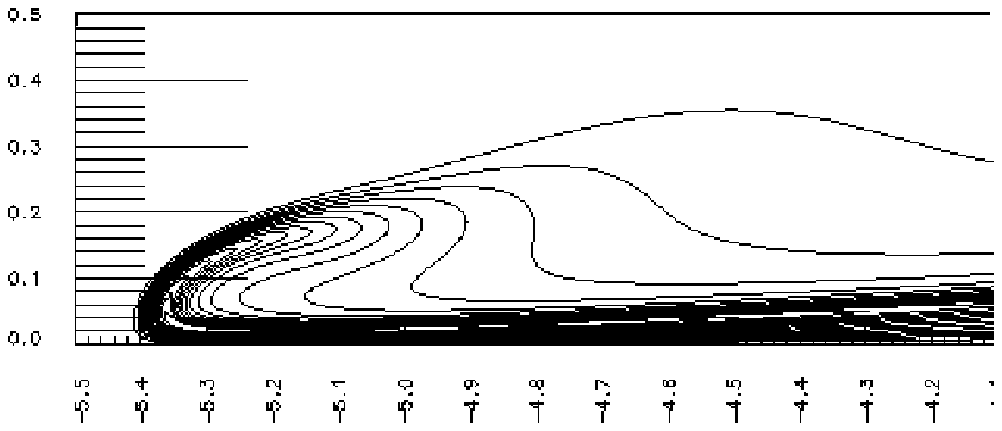}
\includegraphics[width=8cm]{pics/fig4.noscale.t=18.5.epsf}
\includegraphics[width=8cm]{pics/fig4.noscale.t=19.epsf}
\end{center}
\caption{The contour of axial vorticity around the maximum vorticity on
the symmetry plane (the $x-z$ plane) at $t=17.5, 18, 18.5, 19$. 
\label{fig.local_struc}}
\end{figure}

In order to see better the dynamic development of the local vortex structure,
we plot a sequence of vorticity contours on the symmetry plane at 
$t=17.5, 18, 18.5,$ and $19$ respectively in Figure \ref{fig.local_struc}.
The pictures are plotted using the
original length scales, without the scaling by a factor of 5 in the $z$ 
direction as in Figure \ref{fig.scaled_local_struc}. From these results, we 
can see that the vortex sheet is compressed in the $z$ direction. 
It is clear that a thin layer (or a vortex sheet) is formed dynamically. 
The head of the vortex sheet begins to roll up around $t=16$. Here the
head of the vortex sheet refers to the region extending above
the vorticity peak just behind the leading edge of the vortex
sheet \cite{Kerr93}. By the time $t=19$, the head of the vortex sheet 
has traveled backward for quite a distance and away from the dividing
plane. The vortex sheet has been compressed quite strongly along the 
$z$-direction. In order to resolve this nearly singular layer structure,
we use 3072 grid points along the $z$-direction, which gives about
16 grid points across the layer at $t=18$ and about 8 grid points 
across the layer at $t=19$. In comparison, the 192 Chebyshev grid 
points along the $z$-direction in Kerr's computations would not be 
sufficient to resolve the rolled-up portion of the vortex sheet. 

\begin{figure}
\begin{center}
\includegraphics[width=10cm]{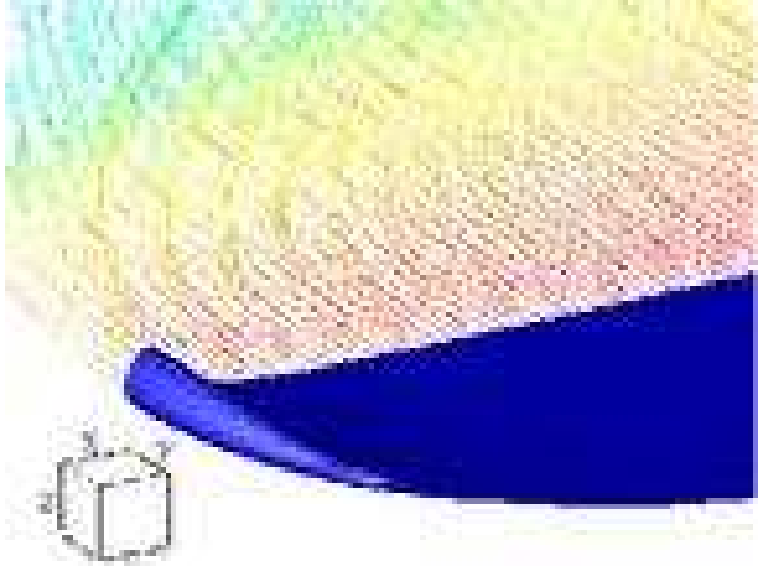}
\end{center}
\caption{The local 3D vortex structure and vortex lines around the maximum
vorticity at $t=17$. The size of the box on the left is 
$0.075^3$ to demonstrate the scale of the picture. 
The isosurface is set at $0.6 \times \|\vec{\omega}\|_\infty$.
\label{fig.local_struc_3d_17}}
\end{figure}

\begin{figure}
\begin{center}
\includegraphics[width=8cm]{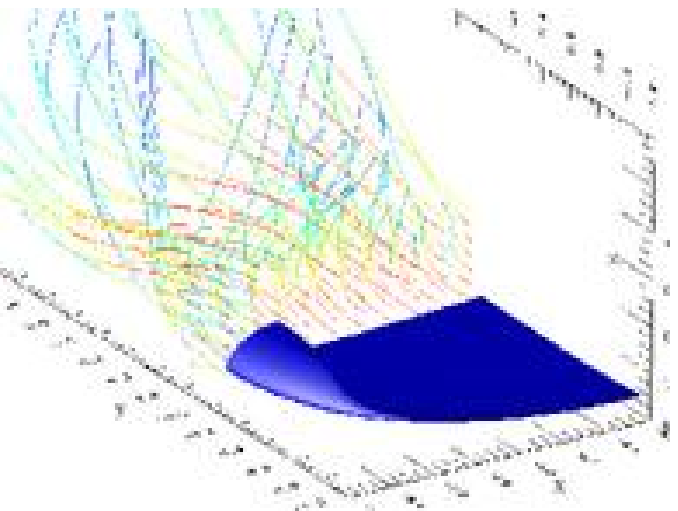}
\hspace{2mm}
\includegraphics[width=8cm]{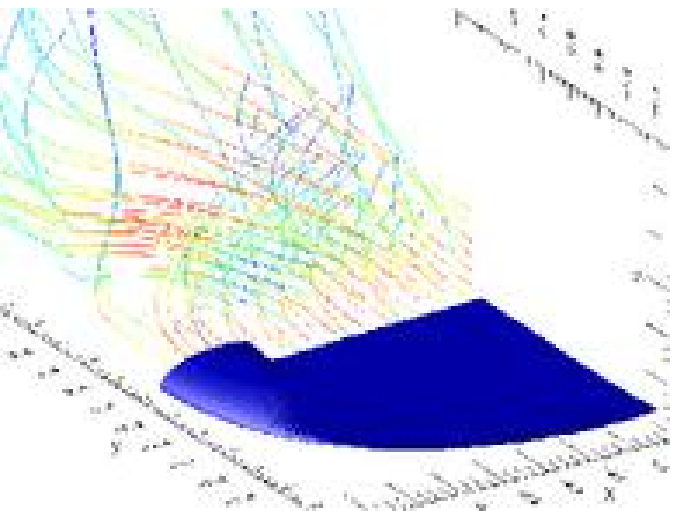}
\end{center}
\caption{
The local 3D vortex structures and vortex lines around the maximum
vorticity at $t=18$ (on the left) and $t=19$ (on the right). 
The isosurface is set at $0.5 \times \|\vec{\omega}\|_\infty$.
\label{fig.local_struc_3d}}
\end{figure}

We also plot the isosurface of vorticity 
near the region of the maximum 
vorticity in Figures \ref{fig.local_struc_3d_17} and \ref{fig.local_struc_3d} 
to illustrate the dynamic roll-up of the vortex sheet near the region of the 
maximum vorticity. The isosurface of vorticity in 
Figure \ref{fig.local_struc_3d_17} is set at 
$0.6 \times \|\vec{\omega}\|_\infty$. 
Figure \ref{fig.local_struc_3d_17} gives the local vorticity structure 
at $t=17$. If we scale the local roll-up region on the left hand side 
next to the box by a factor of 4 along the $z$ direction, as was done 
in \cite{Kerr05}, we would obtain a local roll-up structure which is 
qualitatively similar to Figure 1 in \cite{Kerr05}. 
In Figure \ref{fig.local_struc_3d}, we show the local vorticity structure
for $t=18$ and $t=19$. In both figures, the isosurface
is set at $0.5 \times \|\vec{\omega}\|_\infty$. We can see that
the vortex sheets have rolled up and traveled backward in time
away from the dividing plane. Moreover, we observe that the vortex lines 
near the region of maximum vorticity are relatively straight and the 
unit vorticity vectors seem to be quite regular. On the other hand, the 
inner region containing the maximum vorticity does not seem to shrink 
to zero at a rate of $(T-t)^{1/2}$, as predicted in \cite{Kerr05}. 
The length and the width of the vortex sheet are still $O(1)$, 
although the thickness of the vortex sheet becomes quite small.

Another interesting question is how the vorticity vector aligns with the
eigenvectors of the deformation tensor, which is defined as 
$M \equiv \displaystyle\frac{1}{2} (\nabla \vec{u} + \nabla^T \vec{u}) $.
In Table 1, we document the alignment information of the vorticity 
vector around the point of maximum vorticity with resolution 
$1536\times 1024\times 3072$. In this table, $\lambda_i$ ($i=1,2,3$) is 
the $i$-th eigenvalue of $M$, $\theta_i$ is the angle between the 
$i$-th eigenvector of $M$ and the vorticity vector. One can see clearly
that for $16 \le t \le 19$ the vorticity vector at the point of maximum 
vorticity is almost perfectly aligned with the second eigenvector of $M$.
The angle between the vorticity vector and the second eigenvector is
very small throughout this time interval. Note that the second eigenvalue, 
$\lambda_2$, is positive and is about 20 times smaller in magnitude 
than the largest and the smallest eigenvalues. This dynamic
alignment of the vorticity vector with the second eigenvector of the
deformation tensor is another indication that there is a dynamic 
depletion of vortex stretching.

\begin{table}
\begin{center}
\begin{tabular}{||c|c||c|c||c|c||c|c||} \hline
time & $|\omega|$ & $\lambda_1$ & $\theta_1$ & $\lambda_2$ & $\theta_2$ &  $\lambda_3$ & $\theta_3$ \\ \hline
16.012295 &  5.628002 & -1.508771 & 89.992936 & 0.206199 & 0.007159 & 1.302352 & 89.998852 \\ \hline
16.515890 &  7.016002 & -1.864394 & 89.995940 & 0.232299 & 0.010438 & 1.631355 & 89.990387 \\ \hline
17.013589 &  8.910001 & -2.322629 & 89.998141 & 0.254699 & 0.006815 & 2.066909 & 89.993445 \\ \hline
17.515769 & 11.430017 & -2.630440 & 89.969954 & 0.224305 & 0.085053 & 2.415185 & 89.920433 \\ \hline
18.011609 & 14.890004 & -3.625738 & 89.969613 & 0.257302 & 0.036607 & 3.378515 & 89.979590 \\ \hline
18.516346 & 19.130010 & -4.501348 & 89.966725 & 0.246305 & 0.036617 & 4.274913 & 89.984720 \\ \hline
19.014394 & 23.590012 & -5.477438 & 89.966055 & 0.247906 & 0.034472 & 5.258292 & 89.994005 \\ \hline
\end{tabular}
\end{center}
\caption{The alignment of the vorticity vector and the eigenvectors of $M$
around the point of maximum vorticity with resolution $1536\times 1024\times 3072$.
Here, $\lambda_i$ ($i=1,2,3$) is the $i$-th eigenvalue of $M$, $\theta_i$ is the 
angle between the $i$-th eigenvector of $M$ and the vorticity vector. 
One can see that the vorticity vector is aligned very well with the second 
eigenvector of $M$}. 
\label{tbl.vor_res3}
\end{table}

\section{Concluding Remarks}

We investigate the interaction of two perturbed vortex tubes for the 3D 
Euler equations using Kerr's initial condition \cite{Kerr93}. We use 
both the pseudo-spectral
method with the standard 2/3 dealiasing rule and the pseudo-spectral 
method with a 36th order Fourier smoothing. We perform a careful 
resolution study to demonstrate the convergence of both methods. 
Our numerical computations demonstrate that while both methods converge
to the same solution under resolution study, the pseudo-spectral method
with the 36th order Fourier smoothing offers better computational accuracy
for a given resolution. Moreover, we find that the pseudo-spectral method
with the 36th order Fourier smoothing is more effective in reducing
the numerical oscillations due to the Gibbs phenomenon while still keeping 
the aliasing error under control. 

Our numerical study indicates that 
there is a very subtle dynamic depletion of vortex stretching. The 
maximum vorticity is shown to grow no faster than doubly exponential 
in time up to $T=19$, beyond the singularity time predicted by 
Kerr in \cite{Kerr93}. The velocity field is shown to be bounded 
throughout the computations. Vortex lines near the region of the 
maximum vorticity are quite regular. We provide numerical evidence 
that the vortex stretching term is only weakly nonlinear and is bounded
by $ \|\vec{\omega} \|_\infty \log (\|\vec{\omega} \|_\infty )$. 
This implies that there is tremendous dynamic cancellation in the 
nonlinear vortex stretching term. With the velocity field being bounded
and the vortex lines being regular near the region of the maximum
vorticity, the non-blowup conditions of Deng-Hou-Yu \cite{DHY05a} are 
satisfied. This provides a theoretical support for our computational 
results and sheds some light to our understanding of the dynamic 
depletion of vortex stretching.

Finally, we would like to mention that we have carried out a rigorous
convergence study of the two numerical methods we consider in this 
paper for the one-dimensional Burgers equation. The Burgers equation 
shares some essential numerical difficulties with the the 3D Euler 
equations that we consider here. It has the same type of quadratic 
nonlinearity in the convection term and it is known that it can form 
a shock singularity in a finite time. An important advantage of the 
Burgers equation is that we have an analytic solution formula which can 
be solved numerically up to the machine precision by using the Newton 
iterative method. Using this semi-analytical solution, we have computed 
the solution very close to the shock singularity time and documented 
the errors of both numerical methods using very large resolutions. The 
computational results we obtain on the Burgers equation completely 
support the convergence study of the two numerical methods for the 
3D Euler equations that we present in this paper. The performance 
of these two numerical methods and their convergence property for the
1D Burgers equation are basically the same as those for the 3D 
Euler equations. The detail of this result will be reported elsewhere.

\vspace{0.2in}
\noindent
{\bf Acknowledgments.}
We would like to thank Prof. Lin-Bo Zhang from the Institute of
Computational Mathematics in Chinese Academy of Sciences for 
providing us with the computing resource to perform this large
scale computational project. Additional computing resource was 
provided by the Center of Super Computing Center of Chinese 
Academy of Sciences. We also thank Prof. Robert Kerr for providing 
us with his Fortran subroutine that generates his initial data. 
This work was in part supported by NSF under the NSF FRG grant 
DMS-0353838 and ITR Grant ACI-0204932. Part of this work was done 
while Hou visited the Academy of Systems and Mathematical Sciences 
of CAS in the summer of 2005 as a member of the Oversea Outstanding 
Research Team for Complex Systems. 

\bibliographystyle{amsplain}
\bibliography{bib}

\providecommand{\bysame}{\leavevmode\hbox to3em{\hrulefill}\thinspace}
\providecommand{\MR}{\relax\ifhmode\unskip\space\fi MR }
\providecommand{\MRhref}[2]{%
  \href{http://www.ams.org/mathscinet-getitem?mr=#1}{#2}
}
\providecommand{\href}[2]{#2}
\begin{thebibliography}{10}

\bibitem{AG89}
C.~Anderson and C.~Greengard, \emph{The vortex ring merger problem at infinite
  reynolds number}, Comm. Pure Appl. Maths \textbf{42} (1989), 1123.

\bibitem{BP94}
O.~N. Boratav and R.~B. Pelz, \emph{Direct numerical simulation of transition
  to turbulence from a high-symmetry initial condition}, Phys. Fluids
  \textbf{6} (1994), no.~8, 2757--2784.

\bibitem{BPZ92}
O.~N. Boratav, R.~B. Pelz, and N.~J. Zabusky, \emph{Reconnection in
  orthogonally interacting vortex tubes: Direct numerical simulations and
  quantifications}, Phys. Fluids A \textbf{4} (1992), no.~3, 581--605.

\bibitem{Caf93}
R.~Caflisch, \emph{Singularity formation for complex solutions of the 3{D}
  incompressible {E}uler equations}, Physica D \textbf{67} (1993), 1--18.

\bibitem{Chorin82}
A.~Chorin, \emph{The evolution of a turbulent vortex}, Commun. Math. Phys.
  \textbf{83} (1982), 517.

\bibitem{Const94}
P.~Constantin, \emph{Geometric statistics in turbulence}, SIAM Review
  \textbf{36} (1994), 73.

\bibitem{CFM96}
P.~Constantin, C.~Fefferman, and A.~Majda, \emph{Geometric constraints on
  potentially singular solutions for the 3-{D} {E}uler equation}, Commun. in
  PDEs. \textbf{21} (1996), 559--571.

\bibitem{DHY05a}
J.~Deng, T.~Y. Hou, and X.~Yu, \emph{Geometric properties and non-blowup of
  3-{D} incompressible {E}uler flow}, Comm. in PDEs. \textbf{30} (2005), no.~1,
  225--243.

\bibitem{DHY05b}
\bysame, \emph{Improved geometric conditions for non-blowup of 3{D}
  incompressible {E}uler equation}, Comm. in PDEs. \textbf{31} (2006), no.~2,
  293--306.

\bibitem{FZG95}
V.~M. Fernandez, N.~J. Zabusky, and V.~M. Gryanik, \emph{Vortex intensification
  and collapse of the {L}issajous-{E}lliptic ring: Single and multi-filament
  {B}iot-{S}avart simulations and visiometrics}, J. Fluid Mech. \textbf{299}
  (1995), 289--331.

\bibitem{GMG98}
R.~Grauer, C.~Marliani, and K.~Germaschewski, \emph{Adaptive mesh refinement
  for singular solutions of the incompressible {E}uler equations}, Phys. Rev.
  Lett. \textbf{80} (1998), 19.

\bibitem{GS91}
R.~Grauer and T.~Sideris, \emph{Numerical computation of three dimensional
  incompressible ideal fluids with swirl}, Phys. Rev. Lett. \textbf{67} (1991),
  3511.

\bibitem{HL06}
T.~Y. Hou and R.~Li, \emph{Dynamic depletion of vortex stretching and
  non-blowup of the 3-{D} incompressible {E}uler equations}, accepted by J.
  Nonlinear Science. (2006).

\bibitem{Kerr93}
R.~M. Kerr, \emph{Evidence for a singularity of the three dimensional,
  incompressible {E}uler equations}, Phys. Fluids \textbf{5} (1993), no.~7,
  1725--1746.

\bibitem{Kerr97}
\bysame, \emph{{E}uler singularities and turbulence}, 19th ICTAM Kyoto '96
  (T.~Tatsumi, E.~Watanabe, and T.~Kambe, eds.), Elsevier Science, 1997,
  pp.~57--70.

\bibitem{Kerr99}
\bysame, \emph{The outer regions in singular {E}uler}, Fundamental problematic
  issues in turbulence (Birkh\"auser) (Tsnober and Gyr, eds.), 1999.

\bibitem{Kerr05}
\bysame, \emph{Velocity and scaling of collapsing {E}uler vortices}, Phys.
  Fluids \textbf{17} (2005), 075103--114.

\bibitem{KH89}
R.~M. Kerr and F.~Hussain, \emph{Simulation of vortex reconnection}, Physica D
  \textbf{37} (1989), 474.

\bibitem{MH89}
M.~V. Melander and F.~Hussain, \emph{Cross linking of two antiparallel vortex
  tubes}, Phys. Fluids A (1989), 633--636.

\bibitem{Pelz98}
R.~B. Pelz, \emph{Locally self-similar, finite-time collapse in a high-symmetry
  vortex filament model}, Phys. Rev. E \textbf{55} (1997), no.~2, 1617--1626.

\bibitem{PS90}
A.~Pumir and E.~E. Siggia, \emph{Collapsing solutions to the 3-{D} {E}uler
  equations}, Phys. Fluids A \textbf{2} (1990), 220--241.

\bibitem{SMO93}
M.~J. Shelley, D.~I. Meiron, and S.~A. Orszag, \emph{Dynamical aspects of
  vortex reconnection of perturbed anti-parallel vortex tubes}, J. Fluid Mech.
  \textbf{246} (1993), 613--652.

\end{thebibliography}

\end{document}